\documentclass[10pt, twocolumn]{revtex4-2}
\usepackage{tikz}
\usepackage{amsmath}
\usetikzlibrary{positioning}
\usetikzlibrary{shapes.geometric, arrows}
\tikzstyle{startstop} = [circle, minimum width=3cm, minimum height=1cm,text centered, draw=black, fill=blue!30!white]
\tikzstyle{q} = [circle, minimum width=3cm, minimum height=1cm, text centered, draw=black, fill=blue!40!white]
\tikzstyle{arrow} = [thick,->,>=stealth]
\usepackage{subcaption}
\usepackage{enumitem}
\usepackage{braket}
\usepackage{bbold}
\usepackage{hyperref}

\newlist{subenumerate}{enumerate}{2}
\setlist[subenumerate,1]{label=\arabic{subenumeratei}.\arabic*}
\setlist[subenumerate,2]{label=\arabic{subenumeratei}.\arabic{subenumerateii}}

\setlist[enumerate]{label=\arabic*.}
\newcounter{protocol}

\usepackage{datetime}
\usepackage[T1]{fontenc}
\newdateformat{monthyeardate}{\monthname[\THEMONTH] \THEYEAR}
\usepackage{caption}

\begin{document}
\onecolumngrid
\begingroup  
\centering
\Large \textbf{Hardware requirements for trapped-ion based verifiable blind quantum computing with a measurement-only client}\\[1.5em]
\large J van Dam$^{1,2,3,*}$, G Avis$^{1,2,3,4}$, TzB~Propp$^{1,2,3}$, F~Ferreira~da~Silva$^{1,2,3}$, JA~Slater$^5$, TE~Northup$^6$ and S~Wehner$^{1,2,3}$ \\

\endgroup
\vspace{10pt}
\small \noindent \par
$^1$ QuTech, Delft University of Technology, Lorentzweg 1, 2628 CJ, Delft, The Netherlands\\
$^2$ Kavli Institute of Nanoscience, Delft University of Technology, Lorentzweg 1, 2628 CJ, Delft, The Netherlands\\
$^3$ Quantum Computer Science, EEMCS, Delft University of Technology, Lorentzweg 1, 2628 CJ, Delft, The Netherlands\\
$^4$ College of Information and Computer Science, University of Massachusetts, 140 Governors Dr, MA 01002, Amherst, United States\\
$^5$ Q*Bird, Delftechpark 1, 2628 XJ, Delft, The Netherlands\\
$^6$ Institut für Experimentalphysik, Universität Innsbruck, Technikerstraße 25, 6020 Innsbruck, Austria\\
$^*$ Corresponding author: j.vandam-3@tudelft.nl\\

\begin{abstract}
In blind quantum computing, a user with a simple client device can perform a quantum computation on a remote quantum server such that the server cannot gain knowledge about the computation.
Here, we numerically investigate hardware requirements for verifiable blind quantum computing using an ion trap as server and a distant measurement-only client. While the client has no direct access to quantum-computing resources, it can remotely execute quantum programs on the server by measuring photons emitted by the trapped ion. We introduce a numerical model for trapped-ion quantum devices in NetSquid, a discrete-event simulator for quantum networks. Using this, we determine the minimal hardware requirements on a per-parameter basis to perform the verifiable blind quantum computing protocol. We benchmark these for a five-qubit linear graph state, with which any single-qubit rotation can be performed, where client and server are separated by 50 km. Current state-of-the-art ion traps satisfy the minimal requirements on a per-parameter basis, but all current imperfections combined make it impossible to perform the blind computation securely over 50 km using existing technology. Using a genetic algorithm, we determine the set of hardware parameters that minimises the total improvements required, finding directions along which to improve hardware to reach our threshold error probability that would enable experimental demonstration. In this way, we lay a path for the near-term experimental progress required to realise the implementation of verifiable blind quantum computing over a 50 km distance.

\end{abstract}
\maketitle

\section{Introduction}
Quantum computers may outperform classical computers in a variety of tasks \cite{Feynman1982,shor1994algorithms,grover1997quantum}, but these advantages are so far inaccessible. Moreover, despite progress in realising these devices across a variety of physical platforms \cite{Arute2019,Ebadi2021,Madsen2022,Kim2023,moses2023race}, building and running quantum computers is associated with a large financial cost \cite{WorldEconomicForum,Martin2022,parker2023estimating}. Cloud-based access to quantum servers eliminates the need for users to own large and expensive devices themselves \cite{Soeparno2021} but is unsuitable for use cases involving sensitive data, in which a user requires access to the computational power of the quantum server without revealing the input data, the computation or the output to the owner of the server \cite{Sheng2017,Zhou2021,Li2021}.
\\
Blind quantum computing (BQC) is a technique with which a client can execute quantum algorithms at a remote quantum server without the input, the computation, or its outcome being revealed (apart from an upper bound on the size of the computation) \cite{broadbent2009universal}. Preferably, the client is realised as cheaply as possible, to help make quantum computing more widely available. Two options for realising such a client are for it to have the ability to either send single photons \cite{broadbent2009universal} or measure them \cite{morimae2013blind, fitzsimons2017private}. Alternatively, one can make use of a single-qubit-gate-performing client \cite{li2021blind} or a multi-server approach, in combination with a completely classical client \cite{morimae2013secure, sheng2015deterministic, li2014triple, quan2023verifiable}. In this work, we assume a single-server setup where the quantum capabilities of the client are limited to making measurements. \\
The initial BQC protocol, based on measurement-based quantum computation \cite{raussendorf2000quantum,briegel2009measurement}, was later expanded to include verification to test for correctness; approaches for verification are summarised in reference \cite{gheorghiu2019verification}. In such verifiable blind quantum computing (VBQC) protocols, the client can abort if the outputs of certain tests (either trap based \cite{vbqc}, stabiliser based \cite{hayashi2015verifiable} or classical but introducing computational assumptions \cite{mahadev2018classical}) are not as expected. \\
In realistic near-term noisy quantum devices \cite{Preskill2018quantumcomputingin}, imperfections are inevitable. In verifying tests, imperfections and noise in the system can be mistaken for malicious behaviour of the server, resulting in a computation that will be aborted constantly and the client gaining little information about the operations of the server. There are ways to deal with noise in non-verified protocols \cite{sheng2018blind}, and later a noise-robust verified BQC (rVBQC) protocol was introduced aswell \cite{leichtle2021verifying}. This rVBQC protocol tolerates imperfections from noise or malicious behaviour provided that the server does not fail more than 25\% of verifying tests employed by the client. The robustness to noise is realised by repeating the computation multiple times and performing classical error correction (majority voting) on the results.
Already, there have been proof-of-principle demonstrations of BQC \cite{barz2012demonstration, greganti2016demonstration} and rVBQC \cite{drmota2024verifiable} in laboratory settings. For real-world practicality, however, the client needs to be able to be spatially remote from the server, which will require improvements to existing quantum computing and communication hardware.\\
In this paper, we determine the requirements for performing rVBQC at a metropolitan scale (i.e., the scale of a large city) of 50 km using a trapped-ion-based server and a measurement-only client as depicted in figure \ref{fig:setup}, using the 25\% error tolerance as a threshold. In measurement-based quantum computing, any single-qubit gate can be performed using a five-qubit linear graph state (i.e., five qubits, each in a superposition state, with controlled-$Z$ operations between them) \cite{danos2007measurement}. We use this five-qubit graph state to benchmark the performance of the protocol. We focus on rVBQC because this is feasibly achievable in the near-term.\\
We investigate this implementation of the rVQBC protocol numerically using NetSquid, a discrete event simulator for quantum networks \cite{coopmans2021netsquid}. To this end, we introduce a framework for the modelling of trapped-ion quantum servers, including a NetSquid library \cite{iontrapSnippet}, along with a model of a measurement-only client. The simulation is hardware motivated and takes a set of hardware parameters as input. Using this, we
\begin{enumerate}
    \item Identify the per-parameter minimal requirements for hardware to allow for rVBQC. In each case, we assume perfect performance of all other parameters apart from fibre attenuation. This gives us a strict lower bound for each parameter, which is compared to state-of-the-art performance. We show that current state-of-the-art ion traps satisfy absolute minimal requirements on a per-parameter basis, but all current imperfections combined make it impossible to perform rVBQC securely over 50 km using existing technology. These results are summarised in figure \ref{fig:absmin};
    \item Identify the set of hardware parameters that minimises the total improvement needed over current state-of-the-art parameters to allow for a successful implementation of rVBQC. This reveals which parameters need the most improvement and how far we need to improve them to enable a metropolitan-scale application of rVBQC. We do this by combining our requirement on the error probability with the cost (\cite{da2021optimizing, delfteindhoven}) of a set of hardware parameters into a single-objective minimisation problem. This is fed into a genetic algorithm \cite{Mitchell1996} that minimises the total improvement needed over state-of-the-art performance. These results show there is substantial work left to be done in hardware improvements, and they are summarised in figure \ref{fig:minimp}.
\end{enumerate}
We organise this paper as follows. In section \ref{sec:setup}, we provide details of the physical setup that we simulate, including physical parameters. Then in section \ref{sec:res_dis}, we analyse the two primary results of our numerical analysis discussed briefly above. Section \ref{sec:methods} details the trapped-ion model, the client's measurement apparatus, and the genetic algorithm and cost function that define our optimisation procedure. Finally, in section \ref{sec:future} we discuss possible directions for future work. 

\section{Setup}\label{sec:setup}
\begin{figure*}
    \centering
    \begin{tikzpicture}
\draw[draw=blue!60, fill=blue!5] (0,0) rectangle node[yshift=-2.8cm] {Server} (4, 5) ; 

\draw[draw=blue!60, fill=blue!20] (2,0.833) circle (0.3cm) ; 
\draw[draw=blue!60, fill=blue!20] (2,1.666) circle (0.3cm) ; 
\draw[draw=blue!60, fill=blue!20] (2,2.5) circle (0.3cm) ; 
\draw[draw=blue!60, fill=blue!20] (2,3.333) circle (0.3cm) ; 
\draw[draw=blue!60, fill=blue!20] (2,4.166) circle (0.3cm) ; 
\draw[draw=black!40] (2, 1.133) -- (2, 1.366); 
\draw[draw=black!40] (2, 1.966) -- (2, 2.2); 
\draw[draw=black!40] (2, 2.8) -- (2, 3.033); 
\draw[draw=black!40] (2, 3.633) -- (2, 3.866); 

\draw[draw=blue!60, fill=blue!5] (9, 0) rectangle node[yshift=-2.8cm] {Client} (15.3, 5) ; 

\draw[draw=blue!60, fill=blue!20] (11,0.5) rectangle node[yshift=-1.28cm] {PBS} (13, 2.5) ; 

\draw[draw=blue!60, fill=blue!20] (10.4,0.5) rectangle node[yshift=-1.28cm] {Q} (10.7,2.5) ; 

\draw[draw=blue!60, fill=blue!20] (9.8,0.5) rectangle node[yshift=-1.28cm] {Q} (10.1,2.5) ; 

\draw[draw=blue!60, fill=blue!20] (9.2,0.5) rectangle node[yshift=-1.28cm] {H} (9.5,2.5) ; 

\draw[draw=blue!60, fill=blue!20] (13.5, 0.5) node[yshift=-0.28cm, xshift=0.25cm] {SPD} arc(90:270:-1) --cycle; 

\draw[draw=blue!60, fill=blue!20] (11, 3) -- (13, 3) node [yshift=1.28cm, xshift=-1cm] {SPD} arc(0:180:1) --cycle ; 

\draw[draw=blue!60, fill=blue!20] (9.2, 3) rectangle node[yshift=0.78cm] {CB} (10.8, 4) ; 

\draw (4, 3.5) -- node[yshift=0.22cm, xshift=-0.2cm] {Classical channel} (9.2, 3.5);

\draw (4, 1.5) -- node[yshift=0.22cm, xshift=-2.25cm] {Quantum channel} (13.5, 1.5);

\draw (12, 1.5) -- (12, 3); 

\draw[draw=black!40] (11,0.5) -- (13, 2.5); 

\draw[dotted, draw=blue!70, thick] (12, 4) -- (12, 4.7); 
\draw[dotted, draw=blue!70, thick] (10, 4) -- (10, 4.7) -- (15, 4.7) -- (15, 1.5) -- (14.5, 1.5);

\end{tikzpicture}
    \caption{Setup overview of a trapped-ion quantum server (depicted with 5-qubit linear graph state), connected to a client that measures photons coming in through the quantum channel using half (H) and quarter (Q) waveplates to rotate the qubit state, and measures the qubit using a polarising beam splitter (PBS) and single photon detectors (SPD). The classical channel is used by the client to coordinate with the server. The classical box (CB) is a classical unit that the client uses to choose graph colouring and measurement angles, handle classical communication and perform checks on the results send by the server. The photonic qubits sent out by the server are sent over the quantum channel. The client setup is depicted with more detail to show how this is modelled in the simulation, the exact setup of the server is not shown (i.e., it is kept more general).}
    \label{fig:setup}
\end{figure*}
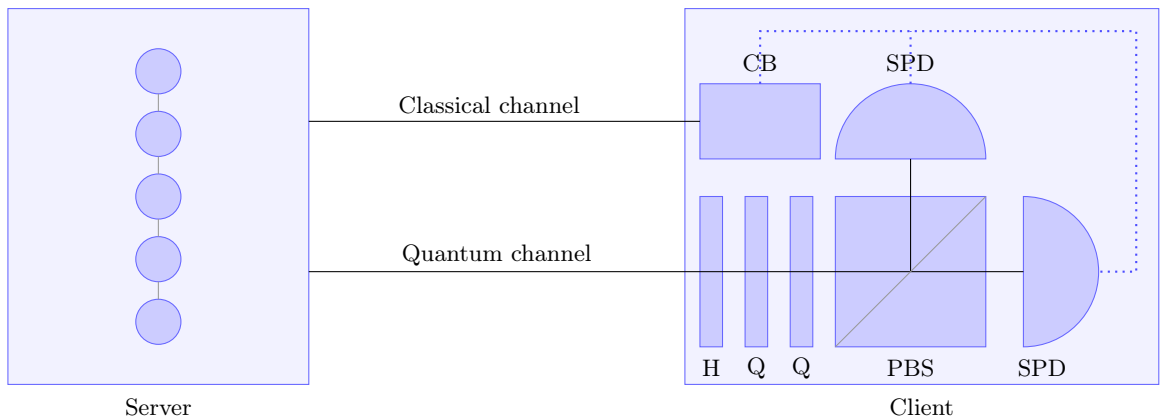

\begin{figure*}
    \centering
    \begin{tikzpicture}
    \definecolor{cblue}{HTML}{7b85d4}
    \definecolor{cred}{HTML}{f37738}
    \definecolor{cgreen}{HTML}{83c995}
    \draw[draw=black!100, fill=black!5] (0, 5.9) rectangle node[yshift=7.4cm] {Client} (6.2, 20.2) ; 
    \draw[draw=black!100, fill=black!5] (10, 5.9) rectangle node[yshift=7.4cm] {Server} (15, 20.2) ; 

    \draw[draw=cblue!100, fill=cblue!40] (0.1, 20.1) rectangle node[] {Choose graph colouring} (6.1, 19.4) ; 
    
    \draw[draw=cblue!100,thick,->] (6.2, 19.4) -- (10, 19.4);
    \draw[draw=cblue!100, fill=cblue!40] (7.6, 19.65) rectangle node[] {I, G} (8.6, 19.2) ; 
    
    \draw[draw=cred!100, fill=cred!40] (10.1, 19.1) rectangle node[] {Initialise ion qubit} (14.9, 18.4) ; 
    \draw[draw=cred!100, fill=cred!40, align=center] (10.1, 18.3) rectangle node[] {Emit ion-entangled \\photonic qubit} (14.9, 17.2) ; 
    
    \draw[draw=cred!100,thick,<-] (6.2, 17.2) -- (10, 17.2);
    \draw[draw=cred!100, fill=cred!40] (6.55, 17.45) rectangle node[] {Photonic qubit} (9.45, 17) ; 
    
    \draw[draw=black!80, fill=black!20, align=center] (0.1, 17.1) rectangle node[] {If dummy} (3, 16.5) ; 
    \draw[draw=black!80, fill=black!20, align=center] (3.1, 17.1) rectangle node[] {If trap} (6.1, 16.5) ; 
    \draw[draw=cred!100, fill=cred!40, align=center] (0.1, 15.3) rectangle node[] {Measure in 0/1} (3, 14.6) ; 
    \draw[draw=cblue!100, fill=cblue!40, align=center] (3.1, 16.4) rectangle node[] {Choose random \\$\theta_v=k_v\pi/4$} (6.1, 15.4) ; 
    \draw[draw=cred!100, fill=cred!40, align=center] (3.1, 15.3) rectangle node[] {Measure in $\pm_{\theta_v}$} (6.1, 14.6) ; 
    \draw[draw=cgreen!100, fill=cgreen!40] (0.1, 14.5) rectangle node[] {Measurement outcome $m_v$} (6.1, 13.8) ; 
    
    \draw[draw=black!100] (-0.2, 19.2) rectangle node[xshift=8cm] {$\times 5$} (15.2, 13.7) ; 

    \draw[draw=cred!100, fill=cred!40, align=center] (10.1, 13.6) rectangle node[] {Prepare graph: \\apply $CZ$ gates} (14.9, 12.5) ; 

    \draw[draw=black!80, fill=black!20, align=center] (0.1, 12.3) rectangle node[] {If dummy} (3, 11.7) ; 
    \draw[draw=black!80, fill=black!20, align=center] (3.1, 12.3) rectangle node[] {If trap} (6.1, 11.7) ; 
    \draw[draw=cblue!100, fill=cblue!40, align=center] (0.1, 11.6) rectangle node[] {Choose random\\ $0\leq n_v \leq 7$} (3, 10.5) ; 
    \draw[draw=cblue!100, fill=cblue!40, align=center] (3.1, 11.6) rectangle node[] {Choose random\\bit $r_v$} (6.1, 10.5) ; 
    \draw[draw=cblue!100, fill=cblue!40, align=center] (0.1, 10.4) rectangle node[] {Set $\delta_v=n_v\pi/4$} (3, 9.3) ;
    \draw[draw=cblue!100, fill=cblue!40, align=center] (3.1, 10.4) rectangle node[] {Set $\delta_v=$\\$\theta_v + m_v \pi + r_v \pi$} (6.1, 9.3) ; 

    \draw[draw=cblue,thick,->] (6.2, 9.3) -- (10, 9.3);
    \draw[draw=cblue!100, fill=cblue!40] (7.7, 9.55) rectangle node[] {$\delta_v$} (8.5, 9.1) ; 

    \draw[draw=cred!100, fill=cred!40, align=center] (10.1, 9.2) rectangle node[] {Measure qubit $v$ in $\delta_v$} (14.9, 8.5) ; 
    \draw[draw=cgreen!100, fill=cgreen!40, align=center] (10.1, 8.4) rectangle node[] {Measurement outcome $b_v$} (14.9, 7.7) ; 

    \draw[draw=cblue!100,thick,<-] (6.2, 7.7) -- (10, 7.7);
    \draw[draw=cblue!100, fill=cblue!40] (7.7, 7.95) rectangle node[] {$b_v$} (8.5, 7.5) ;

    \draw[draw=black!80, fill=black!20, align=center] (3.1, 7.6) rectangle node[] {If trap} (6.1, 6.9) ; 
    \draw[draw=cblue!100, fill=cblue!40, align=center] (3.1, 6.8) rectangle node[] {Check if $b_v=r_v$} (6.1, 6.1) ; 

    \draw[draw=black!80] (-0.2, 12.4) rectangle node[xshift=8cm] {$\times 5$} (15.2, 6.0) ; 
    
\end{tikzpicture}
    \caption{Visualisation of a test round. Depicts tasks (blocks) performed by the client (left) and server (right) along with communication between them (arrows). Blue blocks and arrows represent classical tasks and communication, orange blocks and arrows represent quantum tasks and communication. Grey blocks add comments and green blocks highlight the measurement outcomes. Depicted in chronological order from top to bottom, where the boxed areas (black outline) are repeated five times, assuming they are successful, as these steps are performed for all qubits in the graph.}
    \label{fig:protocol}
\end{figure*}
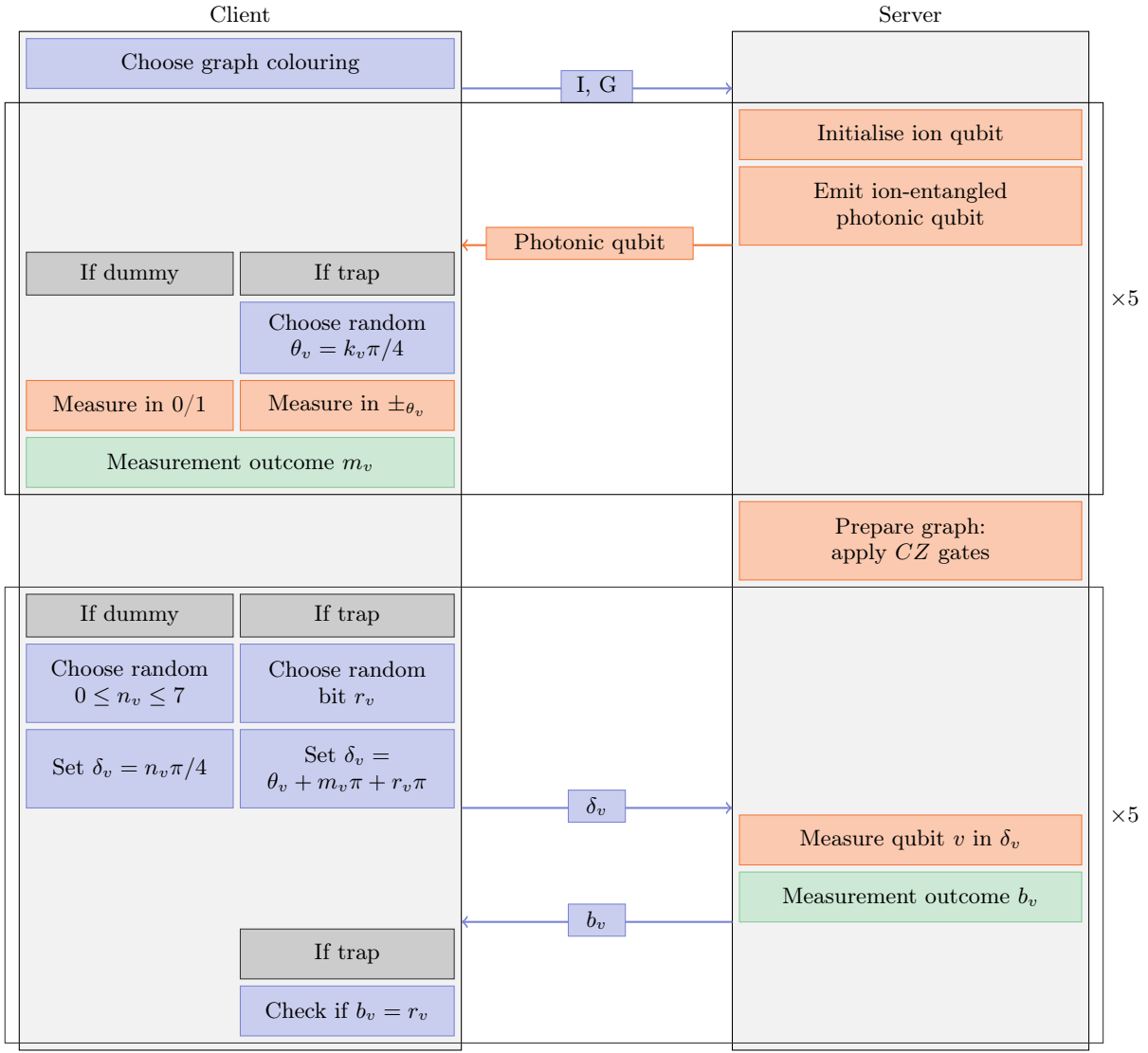

We simulate a two-party rVBQC setup with a trapped-ion quantum server and a measurement-only client using NetSquid. We investigate the protocol at a metropolitan scale, in which the server and client are separated by 50 km of optical fibre. An overview of the setup is provided in figure \ref{fig:setup}. The protocol used here, as in reference \cite{drmota2024verifiable}, assumes a variation wherein the client uses measurements to perform remote state preparation (RSP) on the server. In RSP, a sender measures part of an entangled state and communicates a classical correction to prepare a target state at a receiver.\\
Below, we will outline how normally rVBQC includes both computation and test rounds, and why our analysis only focuses on test rounds (section \ref{subsec:rounds}). In sections \ref{subsec:rsp} through \ref{subsec:output} we describe the steps of such a test round, which summarises the protocol of reference \cite{leichtle2021verifying} and how this is adapted for our simulations. A visualisation of these steps is given in figure \ref{fig:protocol}. After this, we introduce the parameter sets that are used as input to the simulation (section \ref{subsec:params}) and the metrics on which we base our analysis (section \ref{subsec:goals}). This provides necessary context to understand the results as presented in section \ref{sec:res_dis}. \\

\subsection{Computation and test rounds}\label{subsec:rounds}
In a normal run of the protocol, one picks a total number of rounds $N$ that are separated into computation rounds and test rounds, as suggested in reference \cite{leichtle2021verifying}. The output of the total computation is taken as the majority output of the computation rounds. A formal description of the general protocol can be found in \ref{app:formal_protocol}.\\

In our analysis, we focus only on test rounds as the computation rounds do not provide additional information about the verifiability of the setup. From the test rounds, we extract the error probability as the fraction of failed rounds. With the resource we consider, a five-qubit linear graph --- the test rounds can be accompanied by computation rounds in which any single-qubit gate is performed \cite{danos2005parsimonious}. Thus, our results are universal for single-qubit gates.

\subsection{Remote state preparation}\label{subsec:rsp}
The client controls the state of the qubits at the server via RSP. The client can project a qubit located at the server onto a chosen basis by measuring a second qubit, emitted by and entangled with the first qubit (and sent to the client). In this way, the state of the qubits is known to (and determined by) the client, but it is unknown to the server. This is the source of the blindness of this protocol. \\
In each round, the client randomly chooses the sets of qubits that will become the trap qubits and the dummy qubits, such that all trap qubits are surrounded by dummies. For a five-node graph, one can have either dummy-trap-dummy-trap-dummy or trap-dummy-trap-dummy-trap (other combinations are sub-optimal for trap insertion). The dummy qubits need to be prepared in the standard basis $0/1$ (i.e., on the north or south pole of the Bloch sphere), and the trap qubits need to be prepared in a superposition basis $\ket{\pm_{\theta_v}}=(\ket{0}\pm e^{i\theta_v}\ket{1})/\sqrt{2}$ (i.e., on the equator of the Bloch sphere) where $\theta_v \in \{k\pi/4\}_{0\le k \le 7}$, with $k$ an integer such that the angle $\theta_v$ is randomly and independently chosen for each qubit $v$. \\
To start off RSP, the client communicates a description of the graph it wants to prepare to the server. The graph description is the same for all rounds and does not reveal which of the qubits are trap qubits or that a test round is performed at all. This description includes the nodes in the graph $I$ and the edges of the graph $G$ (describing which nodes are connected through $CZ$ gates). Here, we consider a five-qubit linear graph ($I=\{0, 1, 2, 3, 4\}; G=\{(0,1), (1,2), (2,3), (3,4)\}$).\\
After receiving the description, the server emits a polarisation-encoded photonic qubit that is entangled with an ion as $|\psi\rangle = (\ket{0H} + \ket{1V})/\sqrt{2}$, and this photonic qubit is sent to the client over 50 km of optical fibre. The client then sends back a confirmation of arrival to the server and measures the photon in the standard or superposition basis, depending on whether it is a trap qubit or dummy. The client thereby remotely prepares five qubits on the server that are in the state $\ket{m_v}$ or $\ket{+_{\theta_v'}}=(\ket{0} + e^{i\theta_v'}\ket{1})/\sqrt{2}$, where $\theta_v'=\theta_v+m_v\pi$, with $m_v$ the outcome of the client's measurement. Note that the server does not know the basis in which the client measures, and therefore it is unaware of the state of the qubits in its memory. Because the server does not know the state of the qubits, it also cannot tell the difference between a computation round (which involves only $\ket{\pm_{\theta_v}}$-qubits) and a test round (which includes dummy qubits) without performing a (malicious) intermediate measurement that has a chance of disturbing the quantum state.
\\ 
The client performs the superposition basis measurement by rotating the polarisation state of the incoming ion-entangled photonic qubit, then measuring the qubit in the standard basis. Rotating the polarisation state of the photonic qubit can be done by optical elements such as waveplates or electro-optic modulators. A measurement in the standard basis can be performed by a polarising beam splitter (PBS) followed by single photon detectors (SPDs). For a more detailed description of a possible physical realisation of the client, see reference \cite{drmota2024verifiable}. The simulated rotation setup consists of one half waveplate followed by two quarter waveplates, based on reference \cite{simon1990minimal}. 

\subsection{Cutoff time}\label{subesc:cutoff}
In the process of RSP, some qubits will have been prepared and will be sitting in the memory, waiting for the preparation of the remaining qubits. Depending on both server efficiency fibre losses, remotely preparing a qubit might take many tries, meaning that the qubits may have to reside in the memory for a long time. The longer the qubits are in memory, the more they decohere (details in section \ref{subsec:TI}). For a lower error probability, we want to limit the time qubits spend in memory, for which we include a cutoff time.\\
Choosing a good cutoff time is a trade-off between rate and fidelity: a shorter cutoff time gives higher quality qubits as they have suffered from less decoherence, yet more qubits will be discarded, leading to a lower rate. However, since this work only targets an error probability (which depends on the fidelity of the remotely prepared qubits) and does not consider rate, no optimisation can be performed on this aspect. See reference \cite{rozpkedek2018parameter} for examples of optimising this. Instead, we choose a fixed cutoff of half the coherence time.\\ 
When any of the qubits in memory have been at the server for longer than the cutoff time, they are discarded and the client is notified. When the server has prepared five qubits, no more qubits will be discarded (i.e., no cutoff is imposed after RSP).

\subsection{Graph formation}\label{subsec:CZ}
\begin{figure*}
    \centering
    \includegraphics[width=1\textwidth]{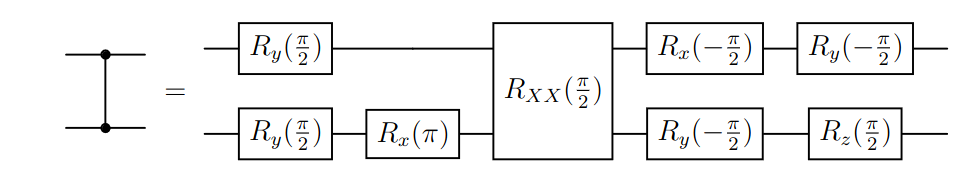}
    \caption{Controlled-$Z$ gate compiled using Qiskit for assumed native gateset of the trapped-ion quantum server. The server is assumed to be able to individually address the ions for rotations ($R_x, R_y, R_z$), and to be able to perform the bichromatic gate ($R_{XX}$, also known as M{\o}lmer-S{\o}rensen gate) as native entangling gate.}
    \label{fig:CZ}
\end{figure*}

The client instructs the server to arrange the qubits in the linear graph: edges should be made between qubits $(0,1), (1,2), (2,3)$ and $(3,4)$. \\
We assume that the ion qubits can be individually addressed for rotations $R_x, R_y, R_z$. This ability is hardware dependent, it is possible in, for example, references \cite{gaebler2016high, pogorelov2021compact, ballance2016high}, each of which uses a different mechanism to implement individual addressing. We note that it is sufficient to implement individual addressing for rotations around just one axis together with collective global rotations around an arbitrary axis in the orthogonal plane \cite{schindler2013quantum}. 
We assume the bichromatic gate $R_{XX}(\theta)=\exp(-i\theta/2 \sum_{k<l}\sigma_x^{(k)}\sigma_x^{(l)})$, also known as the M{\o}lmer-S{\o}rensen gate \cite{sorensen1999quantum, sorensen2000entanglement}, to be the native entangling operation. We use Qiskit \cite{Qiskit} to construct a $CZ$ gate using the operations available for the ion trap (figure~\ref{fig:CZ}). \\
In principle, the server does not need to wait for all qubits to be prepared before it starts applying the $CZ$ operations, but this is done in the simulation for simplicity. The time it takes to prepare the graph state (around 150 $\mu$s) is short compared to the expected time it takes for the qubits to be remotely prepared (on the order of tens of milliseconds\footnote{The average time it takes to remotely prepare a qubit is calculated as time it takes to perform one attempt divided by the average probability of a photon arriving at the client. The average time is calculated as the distance back and forth (send qubit to client and  wait for confirmation) divided by the speed of light in fibre: 2x50 km / ($c/n$) $\approx$ 0.50 ms, with $c$ being the speed of light, and $n \approx 1.5$ being the index of refection of standard telecommunication optical fibre.  The arrival probability is calculated as the server efficiency (given in table \ref{tab:optparams}) times the probability of photon transmitting through the fibre: 0.1325 (from table \ref{tab:optparams}) $\times$ (50km $\times$ 0.2 dB/km)=0.1. This gives  an average probability of photon arriving of 0.01325. Thus, it takes on average 0.50/0.01325=37.8 ms to remotely prepare a qubit.}).

\subsection{Server measurements and output}\label{subsec:output}
Once the graph-state formation is complete, the client instructs the server to measure in  $\pm_{\delta_v}$-bases defined by \cite{leichtle2021verifying}
\begin{align}\label{eq:delta}
        &\delta_v = \theta_v + m_v\pi + r_v \pi, \text{ for trap qubits},\\
        &\delta_v \in \{ k\pi /4\}_{0\leq k\leq 7}, \text{ for dummy qubits}.
\end{align}
Here, $\theta_v$ defines the basis used by the client to perform RSP, and $m_v$ is the outcome of the RSP measurement of the photonic qubit entangled with ion qubit $v$. The random bit $r_v$ is generated by the client to ensure the measurement outcomes appear random. The angles at which the dummy qubits are measured is irrelevant to the client; they just need to be random, so that the server will not be able to identify the qubit as a dummy (which would reveal that the client is executing a test round). \\
The server communicates the outcome of measuring qubit $v$ to the client as $b_v$. For all trap qubits, the client checks if $b_v=r_v$. The test round succeeds if this is true for all trap qubits. If any of the trap qubits do not satisfy $b_v=r_v$, the round fails. The round failing or succeeding is the output for each of the test rounds. 

\subsection{Parameter sets}\label{subsec:params}
The steps from the previous subsections are simulated in NetSquid to find the error probability (i.e., the fraction of failed test rounds) for a given setup. The simulation takes a set of hardware parameters as input. We divide these hardware parameters into two sets: one set that we do not vary during optimisation (parameter set 1), and one set that we do vary during optimisation (parameter set 2). The first set contains parameters that are inherent to the setup (fibre length), parameters that are established optical components with little room for improvement (fibre loss, crosstalk in beamsplitters, waveplate error probability, detector dark-count rate) and the duration of operations at the ion trap. The last is not optimised over as these timescales are already very small compared to the time required to perform RSP (which is in the order of 20 ms per qubit, two orders of magnitude larger than the timescale of performing gates and readout). This first set is presented in table~\ref{tab:steadyparams}. We do not take into account errors due to polarisation drift in fibre. This effect is assumed to be very small due to corrections with techniques such as the fully automated stabilisation using reference pulses described in reference \cite{Treiber2009}.\\
\begin{table}
    \centering
    \begin{tabular}{|p{60mm}|p{18mm}|}
    \hline 
        \textbf{Parameter} & \textbf{Value}\\
        \hline 
        Channel length & 50 km\\
        Photon loss probability in fibre & 0.2 dB/km\\
        Waveplate error probability (fast axis tilt/retardation deviation) & 0.001$^*$\\
        Dark count probability photondetectors & 0.02\%$^{**}$\\
        Crosstalk in polarising beam splitter & 0.0001\\
        Qubit rotation duration & 12 $\mu$s \cite{krutyanskiy2023telecom} \\
        Entangling gate duration & 107 $\mu$s \cite{krutyanskiy2023telecom} \\
        Ion initialisation duration & 300 ns \cite{stephenson2019entanglement}\\
        Photon emission duration & 300 ns \cite{stephenson2019entanglement}\\
        Ion qubit readout duration & 100 $\mu$s \\
        \hline
    \end{tabular}
    \caption{Parameter set 1: Hardware parameters used as simulation input that are not varied over. $^*$ The waveplate error probability is discussed in section \ref{subsec:client}. $^{**}$ The dark count probability is given as $1-e^{-R_{\text{dc}} \tau}$ with $R_{\text{dc}} \approx 1500$ Hz the dark count rate for SPDs such as the SPDMA Si Avalanche Photodetector by Thorlabs and $\tau = 12.5$ ns the detection time window such as in the supplementary material of Reference \cite{drmota2024verifiable}.}\label{tab:steadyparams}
\end{table}
The second set of parameters are the optimisation parameters. These are the parameters we vary in the minimisation methods explained in section~\ref{subsec:minmeth}. In Table~\ref{tab:optparams} we list these parameters as well as what we consider the `baseline'. The baseline reflects the current state of the art in long-distance trapped-ion experiments and is used in the minimisation methods to find the difference between what is currently achievable and what is needed to run the protocol at metropolitan distances, as will be explained in section \ref{subsec:minmeth}. We choose this baseline instead of individual values from current record experiments (such as the long coherence time of reference \cite{wang2017single}, the high gate fidelities of references \cite{ballance2016high, gaebler2016high} or the high entanglement fidelity of reference \cite{bock2018high}) because we believe it is more realistic for long-distance experiments. We are actively looking ahead to what we need to achieve rVBQC at a metropolitan scale (i.e., over a 50 km distance), instead of focusing on in-lab experiments, as rVBQC is already feasible in a lab setting \cite{drmota2024verifiable}.

\begin{table}[t]
    \centering
    \begin{tabular}{|p{48mm}|p{34mm}|}
    \hline 
        \textbf{Parameter} & \textbf{Baseline value}\\
        \hline 
        Server efficiency \newline (= emit $\times$ freq. conversion) & 0.1325 \newline (= 0.53\cite{schupp2021interface} $\times$ 0.25\cite{krutyanskiy2019light})\\
        Single-qubit gate fidelity & 0.99 \cite{krutianskii2022multiqubit}\\
        Entangling gate fidelity & 0.95 \cite{krutianskii2022multiqubit}\\
        Emission fidelity & 0.974 \cite{stute2012tunable} \\
        Coherence time [ms] & 62 \cite{krutyanskiy2023telecom}\\
        \hline
    \end{tabular}
    \caption{Parameter set 2: Baseline of long-distance trapped-ion experiments consistent with the state of the art. We vary over these parameters to find hardware requirements to perform rVBQC. Server efficiency here refers to the total efficiency of preparing an ion-qubit, emitting an ion-entangled photon, coupling to the fibre (combined as the number for `emit' in the table) and converting its frequency to the telecom C band at 1550 nm (`freq. conversion' in the table). The emission fidelity refers to the fidelity of the ion-photon entangled pair when the photon is emitted.}
    \label{tab:optparams}
\end{table}

\begin{figure*}[ht!]
    \centering
    \includegraphics[width=\textwidth]{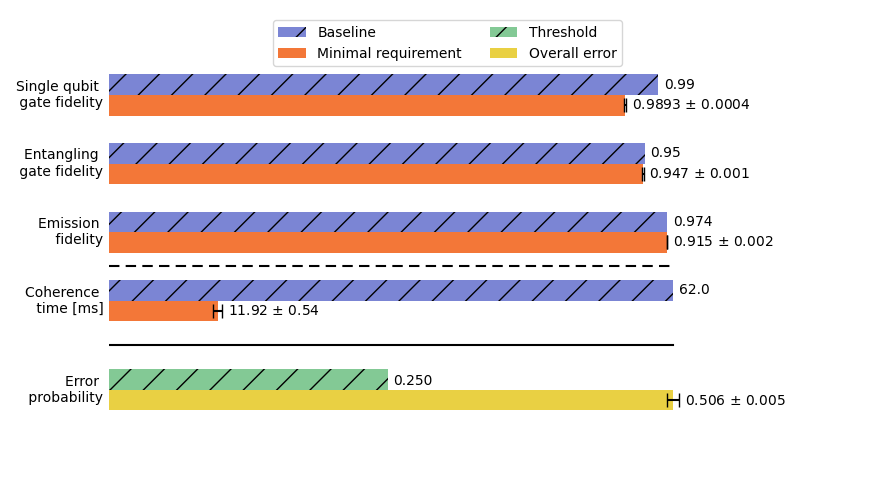}
    \caption{Absolute minimum value a given parameter can have that is consistent with the required maximum error probability of 25\%, assuming that the only other imperfection in the system is photon loss in fibre (orange). These minima are compared to the state of the art as given in Table \ref{tab:optparams} (blue/purple, hatched). The dashed black line indicates a change in scaling between the fidelities and the coherence time. Below the solid black line, the overall error with all imperfections in the state of the art (yellow) is compared with the threshold error probability of 25\% (green hatched); despite each parameter's baseline value meeting the minimal requirements, together the baseline values are insufficient to run the rVBQC protocol.}
    \label{fig:absmin}
\end{figure*}

\subsection{Error tolerance and requirements}\label{subsec:goals}
\section{Results and discussion}\label{sec:res_dis}
\begin{figure*}[!htbp]
  \begin{subfigure}{\textwidth}
    \centering
    \begin{tabular}{|c|c|c|}
        \hline 
         \textbf{Parameter} & \textbf{Baseline} &\textbf{Minimally improved value} \\
         \hline 
         Server efficiency &  0.133 & 0.393\\
         Single-qubit gate fidelity & 0.99 &0.997\\
         Entangling gate fidelity &0.95 & 0.988\\
         Emission fidelity & 0.974&0.982 \\
         Coherence time [ms]&62 & 124 \\
         \hline 
    \end{tabular}
    \caption{}
  \end{subfigure}
  
  \begin{subfigure}{\textwidth}
    \centering
    \includegraphics[width=0.65\linewidth]{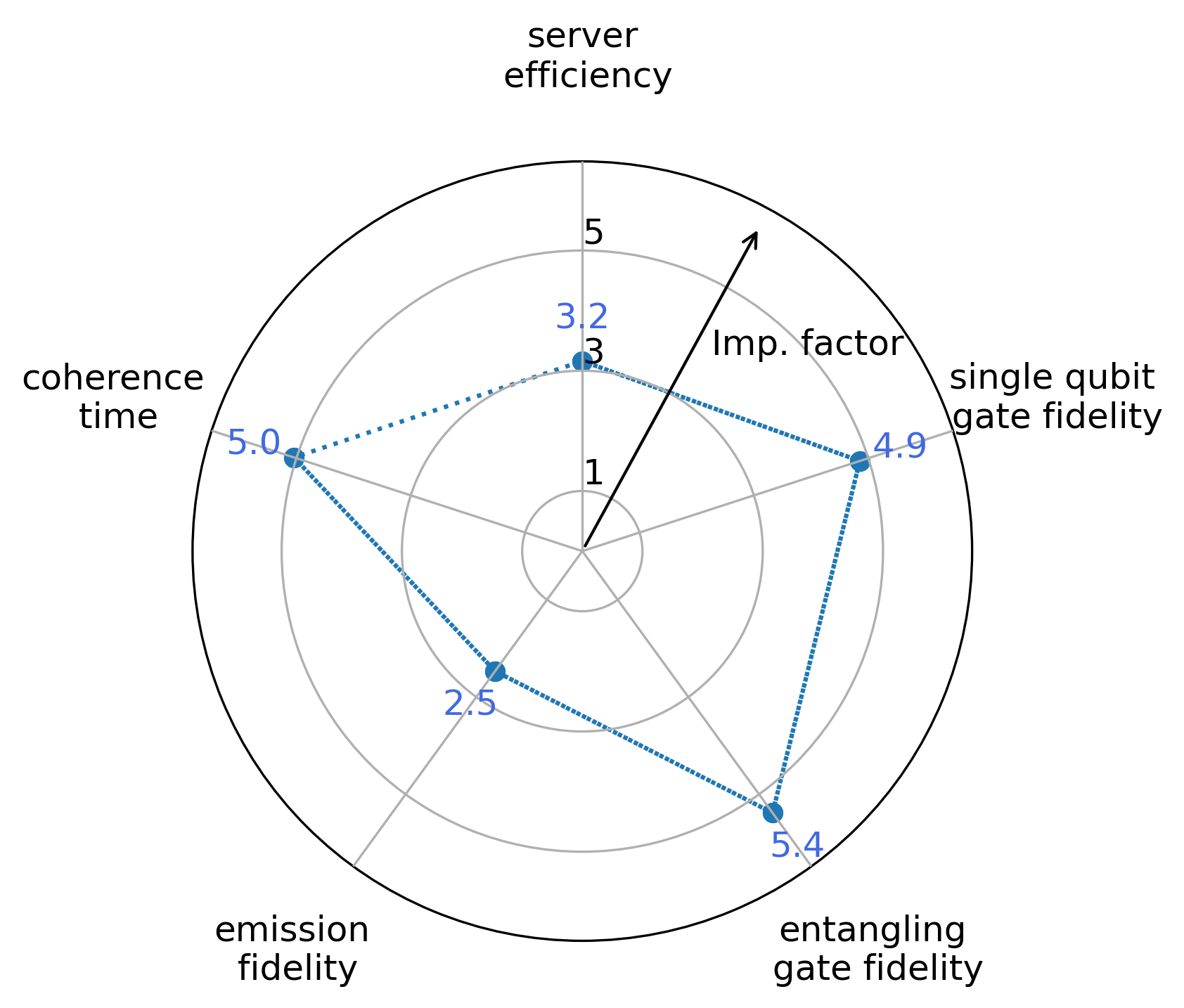} 
    \caption{}
  \end{subfigure}
  
  \caption{\textbf{(a):} Minimally improved parameter set required to perform a 5-qubit linear-graph rVBQC protocol on a trapped-ion server and a measurement-only client over 50 km with a 25\% error probability threshold. This set minimises the cost function (\ref{eq:cosfunction1}), meaning it is the closest to the state-of-the-art baseline of the sets that meet the requirement. \textbf{(b):} Directions along which hardware must be improved to implement a 5-qubit linear graph rVBQC protocol on a trapped-ion server and a measurement-only client over 50 km. The further away the line is from the centre towards a given parameter, the larger the improvement that parameter requires. The improvements are given in terms of an improvement factor $k$, which tends to infinity as a parameter tends to its perfect value and is 1 for no improvement.}\label{fig:minimp}
\end{figure*}
Here, we justify the threshold and explain the two different optimisation procedures we use to come to two different sets of minimised parameters.\\ 
In abstract cryptography (i.e., the theory behind the security proof of the rVBQC protocol \cite{leichtle2021verifying}), security is defined as the indistinguishability between the real-world implementation of the protocol and the ideal, noiseless resource \cite{maurer2016indifferentiability}. When the security of a protocol holds over sequential or parallel repetitions with other protocols, it is said to be \textit{composably secure}. The rVBQC protocol we study (\ref{app:formal_protocol}) is $\epsilon$-composably-secure (as defined in \cite{leichtle2021verifying}) with $\epsilon$ exponentially small in the number of rounds if the fraction of failed test rounds over total number of test rounds is bounded by 25\% \cite{leichtle2021verifying}. Details for this are provided in section \ref{subsec:minmeth}. The protocol includes classical error correction in the form of a repetition code, but includes no quantum error correction. Note that the 25\% error constraint is only present for rVBQC, in non-verifiable BQC the protocol can in principle be performed at any error probability but without any guarantee of the correctness of the outcome. \\
We can use the error tolerance to find two sets of requirements: absolute minimal hardware requirements and minimal improvements. To find the absolute minimal hardware requirements, we start by setting all hardware parameters to perfect except for photon loss in fibre. That is, for parameter set 1, we set the waveplate errors, dark-count probability and beam splitter crosstalk to 0. The loss in fibre is kept as presented in Table~\ref{tab:steadyparams} at 0.2 dB/km. For parameter set 2, we set the server efficiency and all fidelities to 1, and we remove the memory decoherence noise model to simulate effectively infinite coherence time. Then one parameter at a time from set 2 is made progressively worse, until the simulated error probability rises above 25\%. The last value of the parameter before this happens is taken as the absolute minimal requirement. \\
For the minimal improvements, we aim to find the least costly improvements needed over the baseline parameters to get the error probability below 25\%. The cost of a hardware parameter is given in terms of an improvement factor. This improvement factor quantifies the difference between a given value and the baseline, being 1 for no improvement, and tending to infinity for a parameter tending to perfect (e.g., infinite coherence time, fidelity of 1). More information on the cost function and the improvement factor can be found in section \ref{subsec:minmeth}. The minimal improvements are found by combining the error tolerance with the improvement factors into a single objective optimisation problem and solving it using a genetic algorithm. \\
We find requirements for enabling rVBQC with a measurement-only client and a trapped-ion server separated by 50 km of optical fibre, for five-qubit linear graphs.\\
The absolute minimal requirements found, compared to the baseline of Table~\ref{tab:optparams}, are shown in figure \ref{fig:absmin}. We see that state-of-the-art ion traps satisfy absolute minimal requirements on a per-parameter basis. However, we also calculate the error probability with the full baseline of parameter set 1. This shows us that all current imperfections combined make it impossible to perform the blind computation securely over 50 km using existing technology, as the corresponding error probability of 51\% is twice the requirement of 25\%. Note that there is no absolute minimal requirement on the server efficiency, as there is no threshold for rate, only for error probability. If all other parameters are perfect, any server inefficiency combined with fibre attenuation will still lead to a 0\% error probability, as the storage of qubits is perfect. (This is not the case for the minimal improvements set, as other parameters will be imperfect. Having imperfect server efficiency then leads to qubits residing in imperfect memory for longer, thereby suffering more decoherence.) It will only lead to lower rate, not considered here. The difference between the minimal requirement and the state of the art appears particularly large for the coherence time. However, we note that the visualisation is skewed as all other parameters are on a scale of zero to one, which is not the case for coherence time. Because of this we chose to separate the coherence time from other parameters with a dashed line, to indicate a different scaling. \\
We find the minimal improvements, i.e., the set that is closest to the baseline in table~\ref{tab:optparams} satisfying the threshold. The set of hardware parameters that minimises the cost function are given in figure \ref{fig:minimp} (a) and visualised in figure \ref{fig:minimp} (b). The plot in figure \ref{fig:minimp} (b) shows the improvement factor for each parameter in figure \ref{fig:minimp} (a). The further away the line is from the centre towards a given parameter, the larger improvement that parameter requires. From this we can see that comparatively little improvement is needed in terms of improvement factor for server efficiency and emission fidelity. The server efficiency for the ion trap considered here is enhanced due to the use of a cavity \cite{schupp2021interface}.  \\
More improvements are needed for the entangling and single-qubit gate fidelities and for the coherence time. As can be seen from figure \ref{fig:CZ}, many single-qubit gates are executed in order to perform the $CZ$ gate used for creating the graph state, which makes any imperfections in the execution of such single-qubit gates more impactful. We do note that the number of single-qubit gates may be reduced by optimising the graph state creation and perhaps absorbing some of the gates into the measurement bases, which would reduce the overall errors and hence lessen the need for improvement on this parameter. This is not the case for the entangling-gate fidelity.\\
Notably, a lower improvement factor does not always means a value is easier to obtain, this lies in the problem of finding a meaningful cost function. In this case, most values have been obtained in separate optimised experiments, but this is not true for the server efficiency. The baseline used for this optimisation is already optimised for server efficiency as it comes from communication experiments, such that further improvement might be more challenging. Therefore, while the improvement factor might help in visualisation, the true values given in table of figure \ref{fig:minimp}(a) might give a clearer idea of what improvements are needed.\\
Genetic algorithms do not guarantee to find a global minimum, instead several local minima were found, of which the one presented here was the lowest. Other solutions are roughly equivalent but might give slightly more importance on improving one parameter over the other. An alternative solution is given in \ref{app:alt_route} for comparison, and other datasets can be found in \cite{data}. 

\section{Methods}\label{sec:methods}
Here, we discuss some details of how the trapped-ion server and client apparatus are modelled in our NetSquid simulation. We also outline the minimisation method used to determine the requirements identified in the previous section. \\
Though in principle the optimisation can be executed using a different simulator, the choice for using NetSquid as opposed to other quantum network simulators is threefold. First, it is well suited for this type of modelling, as it is a dedicated quantum network simulator that simulates the hardware layer on an appropriate scale. Second, an ecosystem of open-source user-contributed libraries has developed around NetSquid, providing us with useful tools and examples \cite{aeSnippet, nvSnippet}. We have been able to contribute back to this ecosystem by integrating our own library NetSquid-TrappedIons \cite{iontrapSnippet}. Lastly, some of the authors of this work having prior experience with NetSquid made it a natural choice. 

\subsection{Trapped-ion modelling}\label{subsec:TI}
We model the trapped-ion server using the NetSquid-based library NetSquid-TrappedIons~\cite{iontrapSnippet}.
This library was first used in~\cite{delfteindhoven} and is here introduced in more detail.\\
We model the decoherence of trapped-ion qubits over time using a collective Gaussian dephasing process~\cite{zwerger2017}, which can be rewritten as~\cite{delfteindhoven}
\begin{equation}
\rho \to \int_{- \infty}^\infty K_r  \rho K_r^\dagger p(r) dr,
\end{equation}
with
\begin{equation}
K_r =  \exp \left(-i r \frac t \tau \sum_{j=1}^n \sigma_z^{(j)}\right)
\end{equation}
and
\begin{equation}
p(r) = \frac{1}{\sqrt{2\pi}}e^{-r^2/2}.
\end{equation} 
Here, $\sigma_z^{(j)}$ is the Pauli $Z$ operator acting on qubit $j$, $\tau$ is the coherence time of the ion qubit and $t$ is the amount of time that has passed. In addition, $K_r$ is the unitary part of the Kraus operator $K_r \sqrt{p(r) dr}$ satisfying Kraus' theorem: $\int_{- \infty}^\infty K^\dagger_r K_r p(r) dr = \int_{- \infty}^\infty p(r) dr = 1$. By writing the model this way, one makes the following interpretation explicit: all the qubits undergo $Z$ rotations at a constant rate of $2r$ per time interval $\tau$, where $r$ is a random variable with probability distribution $p(r)$. We note that the model is ``collective'' in the sense that there is correlated noise between all the qubits in the same ion trap (they all undergo the same random rotation), and ``Gaussian'' in the sense that the probability that no dephasing error took place decreases with a Gaussian profile over time, which is a consequence of $r$ being normally distributed. The noise process is non-Markovian, which poses a challenge when modelling it in a discrete-event simulator like NetSquid. In NetSquid-TrappedIons, this problem is solved by sampling a value for $r$ from $p(r)$ each time the ion qubit is reinitialised, then evolving the qubits over time using the corresponding unitary operator $K_r$.\\
The emission of entangled photons from an ion qubit is modelled in NetSquid-TrappedIons as the creation of a photon that has a polarisation degree of freedom that is maximally entangled with the state of the ion used in the emission, $\ket{\psi} = (\ket{0H} + \ket{1V})/\sqrt{2}$, followed by the application of a single-qubit depolarising channel on the photon's polarisation.
This results in a Werner state of the form $\frac 3 4 F |\psi\rangle \langle \psi| + \frac{4F-1}{3} \frac{\mathbb 1}{4}$, where $F$ is the fidelity of the state with respect to the perfect state $|\psi \rangle$.

\subsection{Client apparatus modelling}\label{subsec:client}
In simulating the client depicted in figure \ref{fig:setup}, the effect of the waveplates is given by the multiplication of the waveplate Jones matrices with the state vector \cite{jones1941new}. The relative phase retardation induced between the fast axis and the slow axis is $\delta = \pi/2$ for a quarter waveplate and $\delta = \pi$ for a half waveplate. The fast axes of the waveplates also have an angle of $\xi$ radians with respect to the x-axis (which is along the plane of polarisation for linearly polarised light), determining the specific rotation that is implemented. However, errors in the setup can influence the retardation, giving a retardation deviation of $\Delta \delta$. It is also possible to have a deviation in the angle $\xi$ leading to $\Delta \xi$. With this, we can write the Jones matrices in general form \cite{Theocaris1979} to include the errors as 
\begin{align}\label{eq:WPerr}
    &U(\delta^\prime, x^\prime) = e^{-i\delta^\prime /2}\times \notag \\&\begin{pmatrix}
        \cos^2x^\prime + e^{i\delta^\prime}\sin^2x^\prime & (1-e^{i\delta^\prime})\cos x^\prime\sin x^\prime\\ (1-e^{i\delta^\prime})\cos x^\prime\sin x^\prime & \sin^2x^\prime + e^{i\delta^\prime}\cos^2x^\prime
    \end{pmatrix},
\end{align}

where $\delta^\prime = \delta + \Delta \delta$ (with $\delta = \pi/2$ for a QWP, and $\delta = \pi$ for a HWP) and $\xi^\prime = \xi + \Delta \xi$. Together, $\Delta \delta$ and $\Delta \xi$ lead to estimated waveplate error probability as given in table~\ref{tab:steadyparams}. The waveplates are implemented as a custom operation in NetSquid according to the Jones matrix. \\
Reference \cite{simon1990minimal} gives an overview of general qubit rotations in terms of these fast axis settings. We can set the fast axes to correspond to a measurement in the $|\pm_\theta \rangle$-basis as
\begin{align*}
        &\xi_1 = 0;\\
        &\xi_2 = \theta/2; \\
        &\xi_3 = \theta/4 -3\pi/4.
\end{align*}

\subsection{Minimisation methods}\label{subsec:minmeth}
In our analysis, we use one target metric: the error tolerance of 25\%, which is a bound on the fraction of test rounds that are allowed to fail while still being $\epsilon$-composably secure. The exact value of $\epsilon$ is not considered in this analysis, apart from that it can be made exponentially small by increasing the number of test rounds. In reference~\cite{leichtle2021verifying} the fraction of failed test rounds $w$ over the total number of test rounds $t$ is bounded by
\begin{equation}\label{eq:succesbound}
    w/t < \frac{1}{k}\frac{2p-1}{2p-2}.
\end{equation}
Here, $k$ is the principal colouring of the computation graph (which is the smallest number of `colours' or labels one can give to the nodes in a graph such that no two neighbouring nodes have the same colour; see, for example, reference \cite{lewis2015guide}) and $p$ is the inherent error probability of the bounded-error quantum computation. Assuming a computation for which $p=0$ we require the error probability to be below $1/2k$. The one-dimensional graph for single-qubit rotations used in this paper is two-colourable, which gives a maximum error tolerance of 25\%.\\
We look for absolute minimal hardware requirements by setting all but one parameter to perfect aside from fibre attenuation. For the coherence time, perfect means removing the collective dephasing noise model (section \ref{subsec:TI}) from the ions in the trap. We then sweep over the imperfect parameter to find where the error probability due to this imperfection crosses the threshold. \\
To find the crossing point, we do an initial global search with a small number of test rounds per point (to limit computation time) to find the approximate regime in which the error probability would pass the threshold. Once the region is located, the search is focused by taking the closest point above and below the threshold, halving the distance to their mean and running the simulation again for these points with a larger number of rounds. This process is repeated until the error-probability confidence interval of the points crosses the threshold of 25\%. (This is similar to the bisection method used in root finding). The confidence intervals are determined by Hoeffding's bound \cite{hoeffding1965asymptotically} as $\sqrt{\ln{(2/0.05)}/2t}$, with $t$ the number of test rounds. The focused search is executed with 70000 points in order to have a confidence level of 95\% in an interval $\pm$ 0.005 for the error probability. The minimal requirements given in figure \ref{fig:absmin} are then extracted from the closest points ($(x_1, y_1), (x_2, y_2)$) by a linear interpolation at the threshold ($y=0.25$) as $x = (y-y_1)*(x_2-x_1)/(y_2-y_1)$. The error in these estimates is found by applying the same interpolation to the edges of the error probability confidence interval. \\
Next, we find the set of minimal improvements. That is, from a given baseline (table \ref{tab:optparams}), what parameters allows us to fulfill the constraint on the error probability with the least improvement? In order to quantify the cost of improving a parameter by a certain amount, we define hardware costs $H_c$ and a cost function $C$, which combines the hardware costs with our constraint on the error probability to give a single-objective minimisation problem as done in~\cite{da2021optimizing,labay2023reducing,dasilva2023requirements,delfteindhoven}. We then employ a genetic algorithm using a workflow manager called YOTSE \cite{YOTSE} to find the set of hardware parameters that minimises the cost function. This optimisation was run on SURF’s\footnote{SURF is a collaborative organisation for IT in Dutch education and research.} high-performance-computing cluster Snellius (Platinum 8360Y CPU @ 2.4GHz, maximum of 480 GB RAM) and on a workstation featuring an Intel Xeon Gold 6230 CPU @ 2.10GHz and 188 GB of DDR4 RAM, taking around 8000 core-hours per optimisation run of 20 generations.\\
To have a consistent way of calculating hardware cost, we associate a probability of no-imperfection $p_{NI}(b_i)$ to each of the $N$ baseline hardware parameters $b_i\in B$, where $B$ is the baseline set of $N$ hardware parameters. This scales all parameters from 0 to 1, where 1 means a perfect setting (e.g., infinite coherence time, 100\% efficiency). We can improve upon this baseline with an improvement factor $k$ to find $p_{NI}(x_i)=\sqrt[k]{p_{NI}(b_i)}$. These $p_{NI}$ are then summed over for all hardware parameters $x_i$ to find the total hardware cost $H_c(X)$ of a setup with hardware set $X=\{ x_i\}_{0<i\leq N}$ with respect to the baseline $B$ as 
\begin{equation}\label{eq:hardwarecost}
    H_c(X) = \sum_{i=1}^N \frac{\ln \{ p_{NI}(b_i)\}}{\ln \{ p_{NI}(x_i)\}}.
\end{equation}
This is equivalent to summing over the improvement factor of each parameter. The probabilities of no-imperfection are defined for the optimisation parameters as in table~\ref{tab:pnoimp}.\\
\begin{table}
    \centering
    \begin{tabular}{|c|c|}
    \hline 
        \textbf{Parameter} & $\mathbf{p_{NI}}$\\
        \hline 
         Server efficiency $\eta$ & $\eta$\\
         Coherence time $T_c$ & $e^{-t^2/T_c^2}$\\
         (Gate and entangled state) fidelity $F$ & $\frac{1}{3}(4F - 1)$ \\
        \hline
    \end{tabular}
    \caption{Probabilities of no-imperfection: re-scaling parameters from zero to one. This is used to associate an improvement factor from the baseline to each parameter that is dimensionless and thus comparable.}\label{tab:pnoimp}
\end{table}
For derivations and further explanation of the probabilities of no-imperfection, see Supplementary Note 6 of reference \cite{delfteindhoven}. Note that the variable $t$ in the probability of no-imperfection of the coherence time, indicating the timescale over which qubits decohere, does not influence the hardware cost, as 
\begin{equation}
    \frac{\ln \left(p_{NI}(b_i)\right)}{\ln\left( p_{NI}(x_i)\right)} = \frac{\ln\left( e^{-t^2/T_c^2}\right)}{\ln\left( \sqrt[k]{e^{ -t^2/T_c^2}}\right)} = \frac{-t^2/T_c^2}{-t^2/kT_c^2} = k.
\end{equation}
We now combine our requirement of having an error probability below 25\% with the hardware cost to find the total cost of a set of parameters. We want the cost assigned to a set of parameters to be very high when the constraint is not met, and to be lower the closer parameter sets are to the baseline assuming that the constraint is met. A function to capture this behaviour, similar to what is used in references~\cite{da2021optimizing,delfteindhoven,labay2023reducing,dasilva2023requirements}, is 
\begin{equation}\label{eq:cosfunction1}
    C = w_1(1 + (w/t - 1/(2k))^2)\Theta(w/t - 1/(2k)) + w_2 Hc(x_1, ..., x_N),
\end{equation}
where $\Theta(x)$ is the Heaviside step function and $w_1$ and $w_2$ are the weights of the objectives. We choose $w_1 \gg w_2$ in order for the function to reflect that it is much more important to satisfy the error probability requirement than to minimise hardware cost, i.e., we do not care about the hardware cost as long as the requirement is not met. \\
The genetic algorithm is implemented as follows: for all parameters, a number of points are drawn from a range between the baseline (table \ref{tab:optparams}) and their perfect value (except for the coherence time, which is capped at 1 second), i.e., for which $p_{NI}=1$. We initially draw 3 points for server efficiency, 4 for coherence time, 2 for single-qubit fidelity, 3 for entangling gate fidelity and 2 for emission fidelity, meaning the initial population is formed by $3 \times 4 \times 2 \times 3 \times 2 = 144$ sets of parameters. The number of points which are drawn for each parameter is based on the size of the range baseline to perfect for that parameter (e.g., there are more possible values that the entangling gate fidelity can take on than the single-qubit gate fidelity can, as it is further from perfect; we therefore initially draw more points at random from the entangling gate fidelity distribution).
From this initial population, the lowest-costing eight `parents' are taken to recombine with a mutation probability of 0.2 into a new generation. This is repeated over twenty generations. Each point consists of 20000 test rounds for a confidence interval around the error probability of $\pm 0.0096$. The set of parameters with minimal cost is then fed into a local search algorithm, who decreases the cost of each parameter slightly until the error probability requirement is no longer met. The local search is done with 70000 rounds for a confidence interval of $\pm 0.0051$. The outcome of this is a set of parameters that is minimal in the sense that further parameter adjustments that lower the cost of any parameter at that point will result in the requirement not being met.\\
Due to the limits of the search space of the parameters, the lowest value for the hardware cost per parameter is one, when the value of this parameter is equal to that of the baseline. Therefore $H_c$ is bounded from below by 5 (as 5 parameters are considered). The cost does not have an upper bound, and tends to infinity for any parameter tending to its perfect value. In practice, however, the cost does not extend much past $w_1$. 

\section{Future work}\label{sec:future}
One could modify the cost function (\ref{eq:cosfunction1}) to include a constraint on the rate at which the computation or test rounds can be performed. This might change the directions along which the hardware should be improved (i.e., it changes the set of minimal improvements) as parameters such as server efficiency become more important, and conversely the coherence time would become less important. This could be done in the same workflow, by changing the cost function and choosing a rate constraint. We have however chosen not to include this, as there is no immediate clear goal in rate, whereas there is a clear goal in error probability (25\%).\\
The constraint on the error probability of 25\% we consider in this paper is a theoretical limit. In reality, an error probability of 25\% would require an impractically large number of rounds in order to find a desirable $\epsilon$ for security. Instead of setting a constraint on a minimum rate at which the computations can be performed, we might want to find a different metric to target than just the rate and error probability. Having a lower error probability would allow one to perform fewer rounds of the protocol, thereby finishing the total computation faster. An option could be to consider the rate of successfully completed computations, which depends on the error probability, as this determines the number of rounds to be performed, as well as on the rate at which these rounds are performed. This is beyond the scope of this paper. \\
One could also consider requirements for larger graph states. The current analysis considers a universal resource for single-qubit rotations, but a universal resource for any quantum computation (i.e., including two-qubit gates), such as a brickwork state \cite{broadbent2009universal}, would require more qubits. This will lead to more stringent requirements on the hardware. In principle, the same framework used in this paper could be extended to larger graphs such as the brickwork state, but the current state of the code would make the computation time quite a bit longer. We estimate that a 10-qubit graph would take about 2.5 times as long as the 5-qubit graph, the method for this estimation is described in \ref{app:timeest}. This means that the full optimisation procedure (i.e., 20 generations of the genetic algorithm) is estimated to take about 20000 core-hours to complete for a 10-qubit version. The code could be sped up by including a framework similar to NetSquid's entanglement generation Magic \cite{Magicsnippet}, which offers simulation speedup through state insertion. An equivalent to this for RSP is currently in development.\\
The protocol could be optimised by limiting the number of single-qubit gates used in the formation of the graph state, either by considering a full-graph optimisation or by absorbing some of the single-qubit gates into the measurement bases. We also note that, depending on the graph state, not all qubits need to be `alive' in the memory at the same time. The first qubit can be measured before the last qubit is initiated, as long as they are not nearest-neighbours in the graph. In addition, if additional memory qubits are available, the RSP phase could be parallelised by sending ion-entangled photonic qubits from different memory positions successively without waiting for a heralding signal in between. 

\section{Data and code availability}
The code used to generate, process and plot the data is available at \cite{code}, and the data is available at \cite{data}.

\section*{Acknowledgements}
We thank Harold Ollivier and Maxime Garnier for useful discussions on rVBQC. We thank Viktor Krutianskii for useful discussions on trapped-ion devices and for sharing experimental data and parameters. We thank David Maier for providing technical support for the code and execution thereof. We thank Kian van der Enden for critical reading of the manuscript. This project (QIA-Phase 1) has received funding from the European Union’s Horizon Europe research and innovation programme under grant agreement No. 101102140.

\section*{Author contributions}
JvD led the development and execution of the simulation of the protocol, gathered and analysed the results, and prepared this manuscript. GA contributed to development of hardware models of trapped ions and prepared section \ref{subsec:TI} of this manuscript. TzBP and JAS provided feedback and discussions throughout the project. FFdS provided technical support and guidance. TEN helped in the technical understanding of the ion traps and the parameters. All authors revised the manuscript. SW conceived and supervised the project.
\\
\bibliographystyle{unsrt}
\bibliography{bib}
\appendix
\newpage
\onecolumngrid
\section{Formal protocol}\label{app:formal_protocol}
Based on \cite{leichtle2021verifying}, as in \cite{drmota2024verifiable}.\\
\textbf{Clients inputs:} Angles $\{ \phi_v \}_{v \in V}$ for all qubits (\textit{vertices}) $V$, determining the gate(s); a graph $G$; a flow $f$ on $G$, determining the order of measurements.
\\
\textbf{Protocol:}\\
\begin{enumerate}
    \item The client chooses uniformly at random a partition $(C,T)$ of the set of indices of all the rounds in the protocol $N, C\cap T = \emptyset$, with $C$ and $T$ the set of indices for computation and test rounds, respectively. 
    \item For all rounds $n\in N$ the client and server perform the following subprotocol (the client may send a message \textit{redo n} to the server before step 2.3, or the server may send it at any time. Both parties then restart round $n$ with fresh randomness, for more information about this redo feature, see \cite{leichtle2021verifying}):
    \begin{enumerate}[label*=\arabic*.]
        \item If $n\in T$ (test round), the client chooses uniformly at random a colour $K_n$ to define the set of trap vertices for this test round. (The colouring of a graph refers to a way of labelling the nodes such that no neighboring nodes have the same colour, this ensures that no two traps are connected through an edge).
        \item For all $v \in V$ (i.e., for all qubits): \# RSP 
        \begin{enumerate}[label*=\arabic*., align=left, leftmargin=2.5em]
            \item The server prepares a bell pair $|\psi\rangle=(|00\rangle + |11\rangle)/\sqrt{2}$ and sends half of it to the client along with a classical ID.
            \item[2.2.2 a.] When the ID arrives at the client, and the qubit also arrived, the client sends a classical confirmation back to server and performs a measurement yielding outcome $m_v$. The measurement basis is chosen as:
                \begin{enumerate}[label*=\arabic*., align=left]
                    \item[(i)] If $n\in T$ and $v \notin K_n$, measure in the standard basis (i.e. prepare a dummy qubit).
                    \item[(ii)] If $n\in C$ (computation round) or if $n \in T \land v\in K_n$ (trap qubit in test round), measure in $\pm_{\theta_v}$-basis, with $\theta_v\in \{k\pi/4\}_{0\leq k \leq 7}$ randomly chosen,  (i.e. prepare $|\pm_{\theta_v}$).
                \end{enumerate}
            \item[2.2.2 b.] When ID arrives at client, if qubit did not arrive: client sends classical 'lost' message to server. The server removes the qubit from its memory and goes back to step 2.2.1.
        \end{enumerate}
        \item The client sends description of $G$, the server performs a $CZ$ gate between all qubits that share an edge according to $G$ (i.e. the server constructs the graph state).
        \item For all $v\in V$ the client and server perform the following subprotocol:
        \begin{enumerate}[label*=\arabic*., align=left]
            \item the client instructs the server to measure in a $\pm_{\delta_v}$ basis defined by:
        \begin{enumerate}[align=left, leftmargin=2.5em]
            \item[(i)] If $n\in C$: \begin{equation}
                \delta_v = \phi_v' + \theta_v + m_v\pi + r_v\pi,
            \end{equation}
        where $r_v\in_R \{0,1\}$ is chosen uniformly at random. The angle $\phi_v'$ is defined as
        \begin{equation}
            \phi_v' = (-1)^{s_{X,v}}\phi_v + s_{Z,v}\pi,
        \end{equation}
        with
        \begin{equation}\label{eq:measoutcomes_xy}
            s_{X,v} = \bigoplus_{l\in S_{X,v}} s_l, \; s_{Z,v}=\bigoplus_{l\in S_{Z,v}}s_l.
        \end{equation}
         Where $\bigoplus_{l\in S_{X(Z),v}}$ represents a modulo 2 summation over the $X$ ($Z$) dependency set for qubit $v$. The dependency sets are defined by $S_{X,v} = f(v-1)$ and $S_{Z,v} = \{l:v\in N_G(f(l)) \}$, with $N_G$ referring to the neighbors of a qubit, which are all other qubits connected to it through an edge. These adaptations to the measurement angles eliminate the need for bit flip and phase corrections in between the measurements. For more background on how the measurement angles are determined ($\phi_v$) and adapted ($s_{X,v}, s_{Z,v}$), see \cite{danos2007measurement}.
        \item[(ii)] If $n\in T \land v \in K_n$ (trap): 
        \begin{equation}
            \delta_v = \theta_v + m_v\pi + r_v \pi,
        \end{equation}
        i.e. the qubit is being measured in the basis in which it is prepared.
        \item[(iii)] If $n\in T \land v \notin K_n$ (dummy):
        \begin{equation}
            \delta_v \in \{k\pi/4\}_{0\leq k\leq 7},
        \end{equation}
        is randomly chosen.
        \end{enumerate}
        \item The server measures in the basis defined by the client and and sends back the measurement outcome $b_v$.
        \end{enumerate}
        \item For all $\{v: n\in T \land v\in K_n\}$ the client verifies that $b_v=r_v\oplus d_v$, where $d_v=\bigoplus_{i\in N_G(v)}d_i$ is the sum over the measurement outcomes of the neighbouring dummies of qubit $v$. If this is false for any trap qubit in the test round, the test round fails. If the number of test rounds exceeds a certain fraction $w/t$, the client aborts. Here, $w/t<\frac{1}{k}\frac{2p-1}{2p-2}$, introduced in (\ref{eq:succesbound}), is the error threshold to guarantee variability and correctness.
        \item For all $n\in C$, let $y_c$ be the classical output of computation round $c$, the clients checks for a majority output, i.d. checks if there exists a $y$ such that $\left| \{y_c: y_c=y \} \right| > \left| C\right|/2$. If there is a majority output, this $y$ is taken as the protocol output and the client sends an OK to the server. 
    \end{enumerate}
\end{enumerate}
\newpage
\section{Alternative route}\label{app:alt_route}
Multiple sets of parameters can reach the error probability threshold of 25\% for a similar cost. In Section \ref{sec:res_dis} we give a cost-minimised set of parameters (Figure \ref{fig:minimp}) and discuss what what variations in parameters would yield a similar result. The minimised results show less improvement needed in server efficiency and emission fidelity and more in coherence time, entangling and single qubit gate fidelity. How much improvement is needed in these three main objectives varies slightly over different optimisation outcomes. Here, we provide an alternative optimisation outcome compared to the one in the main text to support this observation. In particular, this minimisation puts slightly more emphasis on the single qubit gate fidelity and a little less on the coherence time and entangling gate fidelity compared to the set given in the main text. In addition, the emission fidelity requires more improvement compared to the solution presented in the main text. 

\begin{figure}[h!]
  \begin{subfigure}{\textwidth}
    \centering
    \begin{tabular}{|c|c|c|}
        \hline 
         \textbf{Parameter} & \textbf{Baseline} &\textbf{Minimally improved value} \\
         \hline 
         Server efficiency &  0.133 & 0.594\\
         Coherence time [ms]& 62 & 103 \\
         Single qubit gate fidelity & 0.99 &0.998\\
         Entangling gate fidelity & 0.95 & 0.986\\
         Ion-photon entanglement fidelity & 0.974 &0.988 \\
         \hline 
    \end{tabular}
    \caption{}
  \end{subfigure}
  
  \begin{subfigure}{\textwidth}
    \centering
    \includegraphics[width=0.7\linewidth]{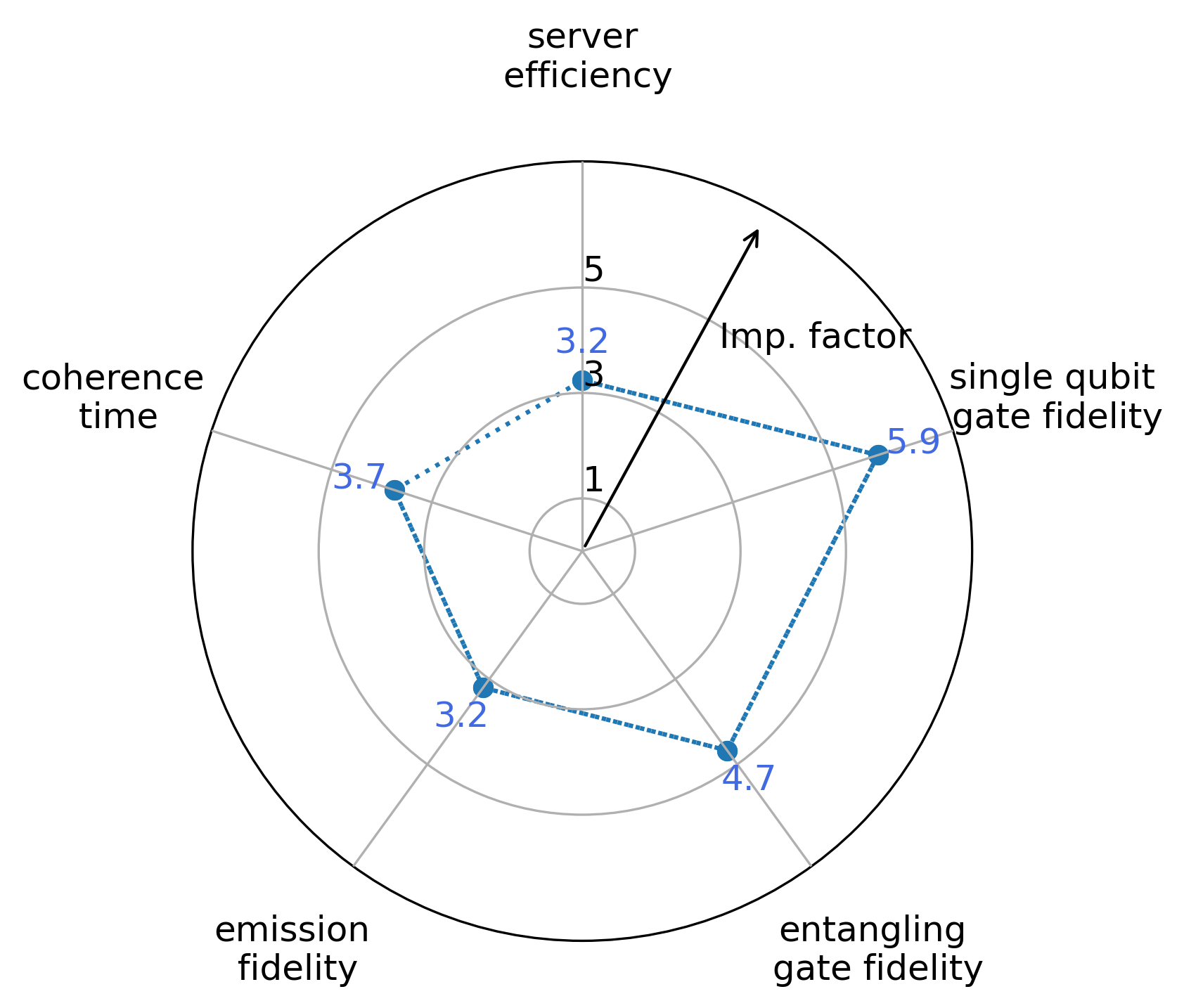} 
    \caption{}
  \end{subfigure}
  
  \caption{\textbf{(a):} Alternative set of parameter values required to perform a 5-qubit linear graph rVBQC protocol on a trapped-ion server and a measurement-only client over 50 km. The requirement for protocol to succeed is having an error probability below 25\%. This set minimises the cost function (\ref{eq:cosfunction1}), meaning it is closest to the state-of-the-art baseline. A visual representation of these parameters in terms of an improvement factor is given in the bottom figure. \textbf{(b):} Directions along which hardware must be improved to perform a 5-qubit linear graph rVBQC protocol on a trapped-ion server and a measurement-only client over 50 km. The further away the line is from the centre towards a given parameter, the larger improvement that parameter requires. The improvements are given in terms of an improvement factor k, which tends to infinity as a parameter tends to its perfect value and is 1 for no improvement. More information on this can be found in Section \ref{subsec:minmeth}.}\label{fig:altminimp}
\end{figure}

\newpage
\section{Estimation of runtime for larger graph states}\label{app:timeest}
\begin{figure}[b]
    \centering
    \includegraphics[width=0.6\linewidth]{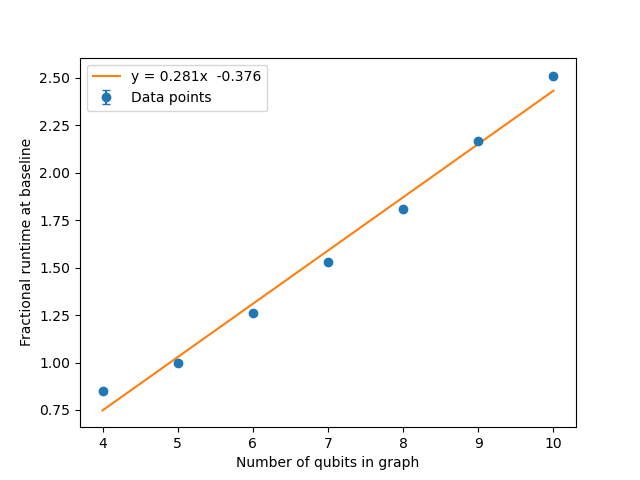}
    \caption{Estimation of the fractional runtime (y-axis) for calculations including a certain number of qubits in the graph (x-axis) compared to the calculation involving a 5-qubit graph. Plot includes standard error of the mean for 5000 points per graph size, error bars smaller than markers (blue). Also plotted with linear fit (orange). This can be used to estimate how long the full optimisation procedure will take.}
    \label{fig:runtime}
\end{figure}

The simulation currently simulates every attempt on remote state preparation; also the failed ones. When the size of the graph gets larger (i.e., it contains more qubits) we will both need to have success more often and, in particular, they all need to happen within the cutoff time, which is half of the coherence time. This will take longer to simulate. The full optimisation procedure (i.e., 20 generations of the genetic algorithm) has not been run for larger graph sizes. To make an estimate of how long this would take, we can compare how long the 5-qubit optimisation procedure took to the fractional runtime of larger graphs at the baseline. When computation B takes twice as long to perform as computation A, the fractional runtime of B compared to A is 2. The optimisation for the 5-qubit graph took roughly 8000 core-hours, a computation with a fractional runtime of 2 compared to the 5-qubit graph would then take roughly 16000 core-hours.\\
To make an estimate of the fractional runtime, we ran the baseline parameters for multiple graph sizes for 5000 test rounds, recorded the time it took to finish and compared this to the time it took to finish the same calculation with a 5-qubit graph. The result of this can be found in figure \ref{fig:runtime}. The fractional runtime seems to increase roughly linearly with increasing number of qubits in the graph, so a linear fit was included to possibly extend this logic to slightly larger graphs. The linear fit results in the relation $y=0.281x - 0.376$ with $x$ the number of qubits in the graph and $y$ the fractional runtime. However, we do not expect this relation to be linear for all graph sizes as we expect that the probability of all qubits being remotely prepared withing the cutoff time will go to zero more quickly for larger graphs. Additionally, this is a rough estimate as it is only fully applicable to the hardware parameters at baseline, whereas the optimisation procedure will have different parameter sets, though we hope it provides a useful estimate nonetheless.

\end{document}